\documentclass[aps,prl,twocolumn,groupedaddress]{revtex4}
\usepackage{epsfig}
\usepackage{latexsym}
\usepackage{graphicx}
\usepackage{bm}

\begin{document}

\bibliographystyle{apsrev}

\preprint{Edinburgh 2007/2, RBRC-638,CU-TP-1174,KEK-TH-1131}

\title{\Large\bf Neutral kaon mixing from 2+1 flavor domain wall QCD}

\newcommand\riken{RIKEN-BNL Research Center, Brookhaven National Laboratory, Upton, NY 11973, USA}
\newcommand\bnlaf{Brookhaven National Laboratory, Upton, NY 11973, USA}
\newcommand\edinb{SUPA, School of Physics, The University of Edinburgh, Edinburgh EH9 3JZ, UK}
\newcommand\epcca{EPCC, School of Physics, The University of Edinburgh, Edinburgh EH9 3JZ, UK}
\newcommand\cuaff{Physics Department, Columbia University, New York, NY 10027, USA}
\newcommand\glasg{SUPA, Department of Physics \& Astronomy, University of Glasgow, Glasgow G12 8QQ, UK}
\newcommand\tokyo{Department of Physics, University of Tokyo, Hongo 7-3-1, Bunkyo-ku, Tokyo 113, Japan}
\newcommand\kanazawa{Institute for Theoretical Physics, Kanazawa University, Kanazawa 920-1192, Japan}
\newcommand\uconn{Physics Department, University of Connecticut, Storrs CT, 06269-3046, USA}
\newcommand\KEK{Institute of Particle and Nuclear Studies, KEK, Tsukuba, Ibaraki 305-0801, Japan}
\newcommand\soton{School of Physics and Astronomy, University of Southampton,  Southampton SO17 1BJ, UK}
\newcommand\boston{Boston University, Boston MA, USA}
\newcommand\tsukuba{Physics Department, SOKENDAI, Tsukuba, Ibaraki 305-0801, Japan}

\newcommand\edmark{$^a$}
\newcommand\rikenmark{$^b$}
\newcommand\cumark{$^c$}
\newcommand\bnlmark{$^d$}
\newcommand\glamark{$^e$}
\newcommand\kanazawamark{$^f$}
\newcommand\sotonmark{$^g$}
\newcommand\uconnmark{$^h$}
\newcommand\KEKmark{$^i$}
\newcommand\bumark{$^j$}
\newcommand\tsukubamark{$^k$}

\author{D.~J.~Antonio\edmark, 
        P.~A.~Boyle\edmark,
        T.~Blum\uconnmark\rikenmark,
        N.~H.~Christ\cumark,
        S.~D.~Cohen\cumark,
        C.~Dawson\rikenmark,
        T.~Izubuchi\rikenmark \kanazawamark,
        R.~D.~Kenway\edmark,
        C.~Jung\bnlmark,
        S.~Li\cumark,
        M.~F.~Lin\cumark,
        R.~D.~Mawhinney\cumark,
        J.~Noaki\sotonmark\KEKmark,
	S.~Ohta\KEKmark\rikenmark\tsukubamark,
        B.~J.~Pendleton\edmark,
        E.~E.~Scholz\bnlmark,
        A.~Soni\bnlmark,
        R.~J.~Tweedie\edmark,
        A.~Yamaguchi\glamark
}
\affiliation{
\centerline{\edmark\edinb}
\centerline{\rikenmark\riken}
\centerline{\cumark\cuaff}
\centerline{\bnlmark\bnlaf}
\centerline{\glamark\glasg}
\centerline{\kanazawamark\kanazawa}
\centerline{\sotonmark\soton}
\centerline{\uconnmark\uconn}
\centerline{\KEKmark\KEK}
\centerline{\tsukubamark\tsukuba}
%\centerline{\bumark\boston}
}

\collaboration{RBC and UKQCD Collaborations : Edinburgh 2007/2, RBRC-638, CU-TP-1174, KEK-TH-1131}

\noaffiliation{RBC and UKQCD Collaborations : Edinburgh 2007/2, RBRC-638, CU-TP-1174, KEK-TH-1131}

\date{October 16, 2007}

\begin{abstract}
We present the first results for neutral kaon mixing using 2+1 flavors
of domain wall fermions.  A new approach is used to extrapolate to the
physical up and down quark masses from our numerical studies with pion 
masses in the range 240 -- 420 MeV; only $SU(2)_L \times SU(2)_R$ 
chiral symmetry is assumed and the kaon is not assumed to be light. Our 
main result is $B_K^{\overline{\rm MS}}(2 \mathrm{~GeV}) = 0.524(10)(28)$ where the 
first error is statistical and the second incorporates estimates for 
all systematic errors.
\end{abstract}

\maketitle

The phenomena of $CP$ violation is a central component of the Standard
Model, in which $CP$ violation is only possible when all three of the quark 
doublets present in Nature interact.  
The Cabibbo-Kobayashi-Maskawa (CKM) flavor mixing matrix contains a single, physically 
meaningful phase which must describe all $CP$ violating phenomena.  For 
bottom mesons, it is possible to make a direct connection between the 
measured $CP$ violation in $B$ decays and this CKM phase.  However, for 
$K$ mesons, the system in which $CP$ violation was originally observed, 
this connection is far more challenging.

One begins with the measure of indirect $CP$ violation 
$\epsilon_K = 2.232 \pm 0.007 \cdot 10^{-3}$ \cite{PDBook}, determined 
experimentally from the mixing between $K^0$ and $\overline{K}^0$ mesons.  
The operator product expansion relates $\epsilon_K$ to the QCD matrix 
element of a four quark operator ${{\cal O}}_{VV+AA} = (\bar{s}\gamma_\mu d) 
(\bar{s}\gamma_\mu d) + (\bar{s}\gamma_5 \gamma_\mu d) 
(\bar{s}\gamma_5 \gamma_\mu d)$  between kaon states via a well known 
perturbative expression \cite{Buras:1998ra} involving this CKM phase.  
This matrix element is parameterized by the renormalization scheme dependent parameter 
\begin{equation}
B_K = \frac{\langle K^0 | {\cal O}_{VV+AA} | \bar{K}^0 \rangle}
                       {\frac{8}{3} f_K^2 M_K^2}.
\end{equation}
Lattice QCD offers the only first-principles determination of $B_K$,
which is essential to determine if the $CP$ violations observed in the 
$B$ and $K$ systems have a common, Standard Model origin.
We describe a lattice QCD calculation of $B_K$ in which the 
most important errors present in earlier lattice results have been 
substantially reduced.  We exploit the domain wall fermion (DWF)
formulation with 2+1 dynamical flavors.  This suppresses $O(a)$ errors,
both on- and off-shell, and also chiral symmetry breaking (measured by a (small) additive ``residual mass''
$m_{\rm res}$). This allows us to renormalize 
${\cal O}_{VV+AA}$ multiplicatively via a non-perturbative 
matching~\cite{Martinelli:1994ty,Blum:2001xb,Blum:2001sr}.  Thus, 
we simultaneously avoid the complexity of lattice operator mixing, 
avoid poorly convergent lattice perturbation theory 
and include the correct light flavor content.  

Alternative lattice approaches to $B_K$ must treat either a chirality 
or taste mixing matrix and result in larger errors.  For
Wilson fermions a chirality mixing matrix can be determined using
non-perturbative off-shell renormalization, but large cancelations
leave results imprecise.  While staggered fermion simulations 
successfully treat simpler quantities, 
current staggered results have a 
10--20\% error for $B_K$ due to large taste mixing \cite{Gamiz:2006sq}.

By using large lattice volumes and meson masses as light as 
243 MeV, we can determine the light quark limit with  substantially 
improved accuracy.  Instead of using a chiral perturbation theory (ChPT) 
which treats the $K$ meson as light compared to the chiral scale,
we evaluate the chiral limit using $SU(2)_L \times SU(2)_R$ ChPT and assume that only our pions are
light.

\section{Simulation}

Our calculation is performed with a fixed lattice spacing and two
space-time volumes, $16^3 \times 32$ and $24^3 \times 64$, using 
the Iwasaki gauge action \cite{Iwasaki:1984cj} with $\beta=2.13$ 
and the DWF action with a fifth dimension of size $16$. 
Each ensemble uses the same dynamical strange quark mass 
$a m_s^{\rm sea} = 0.04$ in lattice units.  We use three $16^3$ ensembles
with degenerate up and down quarks of mass $a m_{l}^{\rm sea} \in \{0.01,0.02,0.03\}$ 
and two $24^3$ ensembles
 with $a m_{l}^{\rm sea} \in \{0.005,0.01\}$.  
The ensembles, described in 
\cite{Allton:2007hx,Antonio:2006xx}, were generated using the RHMC 
algorithm~\cite{Clark:2006fx} with trajectories of unit length.  The 
$16^3$ ensembles each contain 4000 trajectories, from which we omit 
1000 trajectories for thermalization.  We perform measurements on 150 
configurations separated by 20 trajectories for each ensemble, 
calculating matrix elements of all possible pseudoscalar states 
with valence quark masses $a m^{\rm val} \in \{0.01,0.02,0.03,0.04,0.05\}$.  
We bin the data using up to 80 trajectories per bin to reduce the 
correlations between our samples.
The $a m_l^{\rm sea} = 0.005$ and 0.01, $24^3$ ensembles are composed of 4460 and 
5020 trajectories respectively with measurements performed on the 
final 90 configurations separated by 40 trajectories.  These data are 
binned into blocks of 2 configurations representing 80 trajectories.  
All possible pseudoscalar states are studied, composed of valence 
quark masses $a m^{\rm val} \in \{0.001,0.005,0.01,0.02,0.03,0.04\}$.

We use the mass of the $\Omega^-$ baryon, linearly extrapolated to 
$m_l=(m_u+m_d)/2$, and $m_K$ and $m_\pi$ treated in $SU_L(2)\times SU_R(2)$ 
ChPT to determine $1/a=1.73(3)$ GeV, and unrenormalised masses $a m_s + a m_{\rm res} = 0.0375(16)$ 
and $a m_l + a m_{\rm res} = 0.00130(6)$.  Since in our simulation, $a m_s=0.04$ 
and $a m_{\rm res} = 0.00315(2)$, we must take into account our 15\% too 
large input value of $m_s$.  We obtained the pseudoscalar decay constant 
$a f_\pi = 0.0718(18)$ using $Z_A \langle P | A_0 | 0 \rangle= - i M_P f_P$, 
and the axial current renormalization constant $Z_A=0.7162(2)$
\cite{Allton:2007hx}.  (Note, the above errors are all statistical.)

\begin{figure}
\vspace{-0.1in}
\includegraphics[angle=270,width=0.5\textwidth]{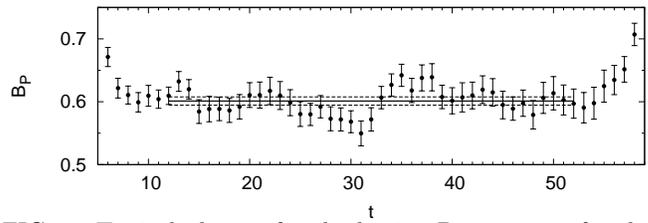}
\vspace{-0.2in}
\caption{\label{figBKplat}Typical plateau for the lattice 
$B$-parameter for the pseudoscalar state made up from quarks of mass 
$a m^{\rm val}_{x}=0.001$, $a m^{\rm val}_y=0.04$,  on the $a m^{\rm sea}_{l}=0.005$, 
$a m^{\rm sea}_s=0.04$, $24^3$ ensemble.
\vspace{-0.1in}
}
\end{figure}

We use an established method \cite{Blum:2001xb,Aoki:2005ga,Aoki:2004ht} 
for evaluating the lattice matrix elements. Zero-momentum kaon states are 
created and annihilated using Coulomb-gauge--fixed wall sources at 
times 5 and 27 for the $16^3$ volume and 5 and 59 for $24^3$, with 
${\cal O}_{VV+AA}$ inserted at each intervening point.  Use of a 
combination of periodic and anti-periodic boundary conditions in time 
removes unwanted propagation around the boundary, resulting in a good 
plateau for the ratio of matrix elements 
\begin{equation}
B_K^{\rm lat} = \frac{\langle K^0(t_1) | {\cal O}_{VV+AA}(t) | \bar{K}^0(t_2) \rangle }
{\frac{8}{3}\langle K^0(t_1) | A_0(t) \rangle \langle A_0(t) | \bar{K}^0(t_2) \rangle}.
\end{equation}
A sample fit is displayed in Fig.~\ref{figBKplat}.  For each ensemble the pseudoscalar mass, $M_P$, decay 
constant, $f_P$, and $B$-parameter, $B_P$, are computed for each combination 
$(m_x^{\rm val},m_y^{\rm val})$. We tabulate a portion of the $16^3$ and 
$24^3$ results for $B_P$ and $M_P$ in Tables~\ref{tabBKvalues16} and 
\ref{tabBKvalues24}.

{
\small
\begin{table}[hbt]
\begin{tabular}{ccccccc}
$a m^{\rm val}$ & \multicolumn{2}{c}{$a m_{l}^{\rm sea}=0.01$} & \multicolumn{2}{c}{$a m_{l}^{\rm sea}=0.02$}
& \multicolumn{2}{c}{$a m_{l}^{\rm sea}=0.03$}\\
 &  $B_P$ & $a M_P$ & $B_P$ & $a M_P$ & $B_P$ & $a M_P$ \\ 
\hline
1,1 & 0.546(8) &0.247(3) &0.539(8) &0.250(3) &0.527(7) &0.251(3) \\
1,2 & 0.577(6) &0.290(3) &0.569(6) &0.292(3) &0.556(6) &0.289(3) \\
2,2 & 0.598(5) &0.323(3) &0.589(5) &0.325(3) &0.580(5) &0.325(3) \\
%1,3 & 0.600(6) &0.325(3) &0.594(6) &0.327(3) &0.579(5) &0.324(3) \\
%2,3 & 0.617(4) &0.357(3) &0.609(5) &0.358(3) &0.600(4) &0.355(3) \\
%3,3 & 0.633(4) &0.385(3) &0.626(4) &0.386(3) &0.618(3) &0.387(3) \\
1,4 & 0.620(5) &0.356(3) &0.616(6) &0.359(3) &0.599(5) &0.356(3) \\
2,4 & 0.633(4) &0.387(3) &0.627(4) &0.388(3) &0.618(4) &0.385(3) \\
3,4 & 0.647(4) &0.414(3) &0.641(4) &0.415(3) &0.634(3) &0.412(3) \\
4,4 & 0.659(3) &0.438(3) &0.655(3) &0.440(3) &0.648(3) &0.442(3) \\
%1,5 & 0.636(5) &0.387(3) &0.634(6) &0.388(3) &0.616(4) &0.386(3) \\
%2,5 & 0.648(4) &0.414(3) &0.643(4) &0.415(3) &0.634(3) &0.413(3) \\
%3,5 & 0.660(3) &0.440(3) &0.655(3) &0.441(2) &0.648(3) &0.438(3) \\
%4,5 & 0.671(3) &0.465(3) &0.667(3) &0.466(2) &0.661(3) &0.463(3) \\
%5,5 & 0.682(3) &0.489(3) &0.679(3) &0.490(2) &0.673(3) &0.488(2) 
\end{tabular}
\caption{\label{tabBKvalues16} Bare pseudoscalar $B$-parameter $B_P$, 
and mass $M_{P}$, for the $16^3$ volume.  The fit range is 12-20 for $
B_P$ and 15-27 for $M_{P}$.   In the first column, {\it e.g.}, 1,2 
denotes a meson composed of quarks with lattice masses 
$(a m_x^{\rm val},a m_y^{\rm val}) = (0.01,0.02)$. 
}
\vspace{-0.2in}
\end{table}

}

{
\small
\begin{table}[hbt]
\begin{tabular}{ccccc}
$a m^{\rm val}_x,$ $a m^{\rm val}_y$
            & \multicolumn{2}{c}{$a m_{l}^{\rm sea}=0.005$} 
                                 & \multicolumn{2}{c}
                                   {$a m_{l}^{\rm sea}=0.01$}\\
            &  $B_P$   & $a M_P$  & $B_P$   & $a M_P$\\ 
\hline
0.001,0.001 & 0.469(8) &0.1402(9) &0.470(5) &0.1434(10) \\
0.001,0.005 & 0.491(7) &0.1681(8) &0.495(4) &0.1707(9) \\
0.005,0.005 & 0.508(5) &0.1916(8) &0.512(3) &0.1938(8) \\
0.001,0.01  & 0.514(6) &0.1971(8) &0.521(4) &0.1995(9) \\
0.005,0.01  & 0.527(4) &0.2172(8) &0.531(3) &0.2194(8) \\
0.01, 0.01  & 0.542(3) &0.2400(7) &0.546(3) &0.2421(8) \\

0.001,0.04  & 0.601(6) &0.3204(11) &0.613(7) &0.3234(11) \\
0.005,0.04  & 0.607(4) &0.3329(8) &0.611(3) &0.3358(8) \\
0.01, 0.04  & 0.614(3) &0.3482(7) &0.616(2) &0.3509(7) \\
\end{tabular}
\caption{\label{tabBKvalues24} Bare pseudoscalar 
$B$-parameter $B_P$, and mass $M_{P}$, results for the 
$24^3$ configurations.  The fit range is 12-52 for the 
$B$-parameter and 15-59 for the mass.}
\end{table}
}

Our lightest dynamical pion masses are 331 MeV. 
We must extrapolate our result for $B_P$ to the physical value 
of $m_l = (m_u+m_d)/2$, and we treat only the up and down quarks
as light by using the $SU(2)_L\times SU(2)_R$ partially quenched ChPT (PQChPT) 
formula \cite{Sharpe:1995qp}
$$B_P(m_x,m_l) = B_P(0)\left\{1 + c_0 m_l + c_1 m_x 
- k \ln \frac{2 B_0 m_x}{\Lambda_\mathrm{ch}^2}\right\}.$$ 
Here $m_x$ is the light valence quark mass, $m_l$ the light sea quark mass,
$\Lambda_\mathrm{ch}$ the chiral scale, $B_P(0)$, $c_0$ and $c_1$ are $m_s$-dependent 
low energy constants, $k=B_0 m_l/(4\pi f_\pi)^2$, $B_0$ is the constant in the expression 
$m_\pi^2 = 2 B_0 m_x$, and we include $m_{\rm res}$ in all masses entering
these formulae.  
%
%%%%%%%%%%%%%%%%%%%%%%% V51 & V52
$m_K$ is now the lowest scale that can dimensionally balance 
higher order terms in the chiral expansion, and $m_\pi^2/m_K^2$ will determine
the suppression of successive orders.
This is a better expansion parameter than the $m_K^2/\Lambda_{\rm ch}^2$
of SU(3) ChPT.
%
%%%%%%%%%%%%%%%%%%%%%%% V50 & 53
%Higher order terms in the chiral expansion in $m_\pi^2/m_K^2$ and $m_\pi^2/(4 \pi f)^2$ 
%are truncated; $m_\pi^2/m_K^2$ represents a better expansion parameter than the $m_K^2/(4\pi f^2)$
%relevant to SU(3) ChPT.
%
%%%%%%%%%%%%%%%%%%%%%%%%%% V49
%%While this treatment 
%%neglects the ratio $(m_l/m_s)^2 \le 12\%$ for $m_l \le 0.01$, such terms, 
%%were they large, might be noticed as causing poor agreement with NLO 
%%$SU(2)_L\times SU(2)_R$ PQChPT.
We should note that the pseudoscalar
masses and decay constants are also well described by a 
$SU(2)_L \times SU(2)_R$ PQChPT analysis.

\begin{figure}
\includegraphics[width=0.4\textwidth]{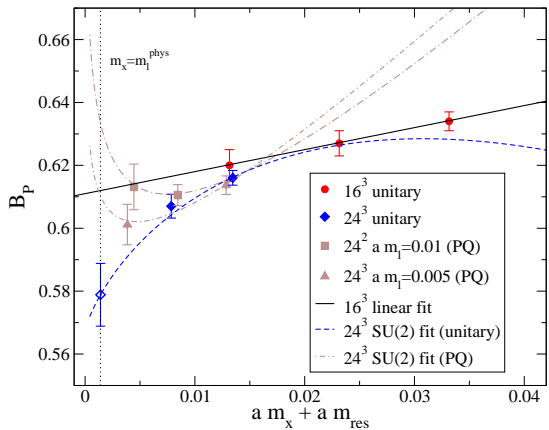}
\caption{\label{fig:chiral}
Results for $B_P$ together with the NLO partially quenched 
$SU(2)_L\times SU(2)_R$ ChPT fit to the $24^3$ data plotted versus the light valence quark 
lattice mass $a m_x$.  From top to bottom on the left-hand-side, the three curves 
are $a m_l$ = 0.01, 0.005 and $a m_x$ respectively.  The valence strange 
quark mass is fixed at its unitary value $a m_y = a m_s = 0.04$.  While 
the statistical errors are large, the growing upward curvature in 
$m_x$ as the sea quark mass is increased from 0.005 to 0.01 
predicted by ChPT is visible.  Some $m_x$ values are slightly 
shifted for clarity.}
\vspace{-0.2in}
\end{figure}

Figure~\ref{fig:chiral} shows $B_P$ versus $m_x$ together with the 
$SU(2)_L\times SU(2)_R$ partially quenched ChPT fit to the $24^3$ data.  The 
fit does not include correlations so the resulting $\chi^2/\mathrm{dof} 
= 0.14$ is not a meaningful indication of goodness of fit.  
The $16^3$ unitary data, also shown in Fig.~\ref{fig:chiral}, are 
well described by a straight line which, if simply extrapolated to
the physical limit gives a result about 6\% larger than the more
accurate chiral extrapolation that is possible if the smaller 
masses in the $24^3$ simulation are used~\footnote{In the original
preprint of this paper, which did not include the $24^3$ results, 
this higher value was obtained.}.  This $SU(2)_L\times SU(2)_R$ chiral
extrapolation gives $B_K^\mathrm{lat} = 0.565(10)$ for physical
$m_l$.  ($B_K$ is determined at the correct valence
value of $a m_s = 0.0343$ by linearly interpolating between $a m_y=0.03$ 
and 0.04.)

%\section{Non-perturbative renormalization}

We have previously demonstrated \cite{Aoki:2005ga,Allton:2006po,
Christ:2005xh} that the wrong-chirality mixing in our simulation
is sufficiently suppressed that we may ignore it.  Thus, we 
consider only multiplicative renormalization of ${\cal O}_{VV+AA}$, 
and use the RI-MOM non-perturbative renormalization technique to 
match our lattice scheme to the $\overline{\rm MS}$ scheme via
$
B_K^{\rm \overline{\rm MS}} =
(Z^{\rm \overline{\rm MS}}_{{\cal O}_{VV+AA}}/Z_A^2) B_K^{\rm lat} 
\equiv Z^{\rm \overline{MS}}_{B_K} B_K^{\rm lat}$.

The technique performs well since domain wall fermions are 
off-shell improved.  We evaluate the amputated, four-leg 
and two-leg vertex functions $\Lambda_{{\cal O}_{\rm VV+AA}}$, 
$\Lambda_A$, and $\Lambda_V$ in Landau gauge.  Because of the
relatively low lattice cut-off, $\Lambda_A$ and $\Lambda_V$ 
differ by 2\%.  We use their average and add their difference
to the systematic error.

\begin{figure}
\vspace{-0.2in}
\includegraphics[width=0.5\textwidth]{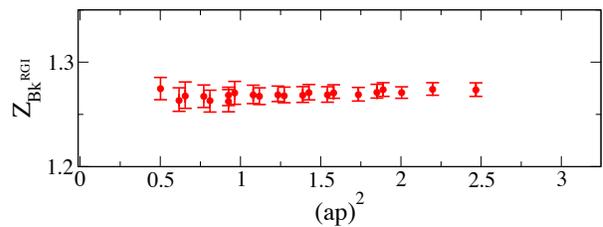}
\vspace{-0.2in}
\caption{\label{figZbk} A plot of $Z_{B_K}^\mathrm{RGI}(p^2)$ showing 
that the perturbative running, removed from $Z_{B_K}^\mathrm{RGI}$, 
accounts for most $p^2$ dependence.}
\end{figure}

We both quote lattice results in the RI-MOM scheme without perturbative 
error and convert to other schemes using the continuum NLO 
result~\cite{Herrlich:1996vf,Ciuchini:1997bw}.   We obtain 
$Z_{B_K}^{\rm RI}(2 \mbox{ GeV}) = 0.910(5)(13)$.  The first 
error in parenthesis is statistical and the second systematic.
As shown in Figure~\ref{figZbk}, we use the $\rm RGI$ scheme as 
a scale invariant intermediate step to reveal and remove 
possible $(ap)^2$ errors.  Only weak scale dependence is seen 
in the window $1.0\le (ap)^2 \le 2.5$, implying artefact-free 
perturbative behavior.  We obtain $Z_{B_K}^{\rm RGI} = 1.275(10)(25)$ 
by linearly extrapolating to $(ap)^2 = 0$ to remove $O(a^2)$ effects.
    
Conversion to $\overline{\rm MS}$ is a 2\% effect at NLO, consistent 
with $O(1)\times \alpha_s/4\pi$.  While the error estimate could be as 
low as $O({\rm few}) \times (\alpha_s/4\pi)^2$, we add in quadrature 
the size of the NLO correction itself as a perturbative systematic, 
giving $Z_{B_K}^{\overline{\rm MS}}(2 \mbox{ GeV}) = 0.928(5)(23)$.
For comparison, 1-loop lattice perturbation theory \cite{Aoki:2002iq} 
gives $Z_{B_K}^{\overline{\rm MS}}(2 \mbox{ GeV}) = 1.007$ with the 
difference likely due to slow convergence of lattice perturbation 
theory. 

\section{Results and Conclusions}

We combine the bare $B_K$ given above with these $Z$ factors
to obtain physically normalized results from
2+1 flavor DWF at $a^{-1} = 1.73(3)$ GeV on a $24^3 \times 64 $ volume,
where the second error is the renormalisation systematic:
$B_K^{\rm RI}(2 \mbox{ GeV}) = 0.514(10)(7)$, and 
$B_K^{\overline{\rm MS}}(2 \mbox{ GeV}) = 0.524(10)(13)$. The RI result
involves no use of perturbation theory and has a reduced systematic error.

\begin{figure}
\vspace{0.1in}
\includegraphics*[width=0.4\textwidth]{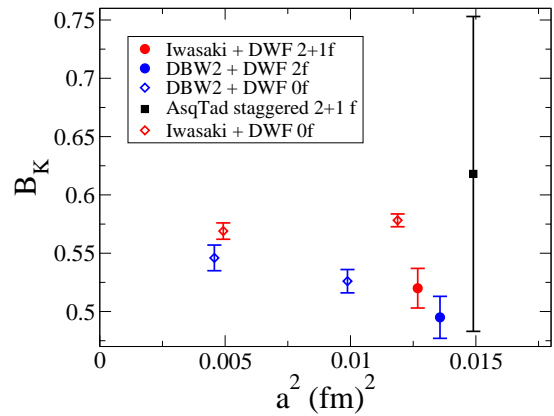}
\caption{\label{figBKworld}
We compare our 2+1 flavor results with earlier quenched\cite{AliKhan:2001wr} and 2 flavor
DWF\cite{Aoki:2004ht} as well as 2+1 flavor staggered calculations\cite{Gamiz:2006sq}.  The quenched Iwasaki
points show statistical errors only while our point and the staggered point include 
renormalisation systematics. While our point lies below these 
Iwasaki results due to our improved chiral limit and flavor content, 
we expect similar $a^2$ dependence.}
\vspace{-0.2in}
\end{figure}

We plot the $\overline{\rm MS}$ result in relation to those quenched\cite{AliKhan:2001wr,
Aoki:2005ga} and two flavor\cite{Aoki:2004ht} DWF results that could 
allow dependence on flavor content to be seen. (Of course differences 
arising from different lattice spacing and gauge action are also to 
be expected.) In addition, we include the 2+1 flavor staggered 
fermion result\cite{Gamiz:2006sq} in Fig.~\ref{figBKworld}.

Finite volume chiral perturbation theory \cite{Becirevic:2003wk} suggests 
that finite volume effects are negligible for all masses and volumes 
in our simulation except for $a m_x=0.001$ where the effect may be 2\%.  
However, since our fit is insensitive to this point, we adopt our
1\%, $16^3$ - $24^3$, $a m_l=0.01$ difference as an estimate of the finite volume 
error.

Finite lattice spacing errors are likely larger, but also difficult for us 
to estimate.  We can make use of the quenched, perturbatively renormalized 
results of CP-PACS\cite{AliKhan:2001wr} also obtained for the Iwasaki and DWF 
lattice action which suggest a (poorly determined) scaling violation of size 
3.5\% at our coarser lattice spacing.  We choose a 4\% systematic error as 
the most likely estimate for $O(a^2)$ effects in our 2+1 flavor result.  A 
second approach to estimating discretization errors is to compare a variety 
of presumably reliable quantities computed from our $1/a=1.73(3)$ GeV 
ensembles with experiment.  For example, we find $f_\pi$ and $f_K$ about 
4\% below experiment, a discrepancy consistent with our 4\% error estimate.

While we have interpolated to the physical valence strange quark
mass, we have results for only one strange sea quark mass.  We 
estimate the error resulting from our 15\% too large $m_s$ by observing 
that for fixed valence masses (0.01,0.04) $B_P$ increases by 3\% when the 
light sea quark mass is changed from 0.03 to 0.02.  Scaling this to the 
0.0057 change needed for a single flavor of sea quark implies a $1\%$ 
error.  Finally we add a 2\% chiral extrapolation error by estimating the 
effects of NNLO ChPT as the 6\% difference between the linear and NLO chiral 
limits in Fig.~\ref{fig:chiral} scaled by $m_l/m_s = 0.4$ for $a m_l=0.01$.

Thus, we take our central value, which removes all quenching systematics, 
and add the 1\% finite volume, 4\% scaling, 1\% $m_s$ extrapolation, 
2\% ChPT and renormalization error estimates in quadrature and obtain:
\begin{eqnarray}
B_K^{\rm RI}(2 \mbox{ GeV})              &=& 0.514(10)(25), \\
B_K^{\overline{\rm MS}}(2 \mbox{ GeV})  &=& 0.524(10)(28), \\
\hat{B}_K                               &=& 0.720(13)(37),
\end{eqnarray}
where the first error is statistical and the second is the estimated 
systematic.  A recent review, including all lattice data then available,
quoted a continuum limit value of $B_K^{\overline{\rm MS}}(2 \mbox{ GeV}) 
= 0.58(3)(6)$\cite{Dawson:2005za}.  Our result is consistent with this 
and reduces both types of error substantially. This improvement arises 
because, using QCDOC computers, we have for the first time combined the 
correct dynamical flavor content with a lattice formulation with good 
chiral symmetry, $O(a)$ improvement, control over operator mixing,
non-perturbative renormalization and a new use of $SU(2)_L\times SU(2)_R$ ChPT
for kaons.

In order to reduce the significant discretisation systematic error in
our result for $B_K$, the RBC and UKQCD collaborations are now doing 
simulations at a smaller lattice spacing, which will give quantitative 
control of this effect. A continuum NNLO perturbative calculation is 
required to convert lattice results to ${\overline{\rm MS}}$ with 
better precision.

We thank Dong Chen,  Mike Clark, Calin Cristian, Zhihua Dong, Alan Gara, 
Andrew Jackson, Changhoan Kim, Ludmila Levkova, Xiaodong Liao, Guofeng
Liu, Konstantin Petrov and Tilo Wettig for developing with us the QCDOC
machine and its software. This development and the computers used in 
this calculation were funded by the U.S.\ DOE grant DE-FG02-92ER40699, 
PPARC JIF grant PPA/J/S/1998/00756 and by RIKEN. This work was supported 
by DOE grant DE-FG02-92ER40699 and PPARC grants PPA/G/O/2002/00465 and 
PP/D000238/1. We thank the University of Edinburgh, PPARC, RIKEN, BNL 
and the U.S.\ DOE for providing these facilities.  C.J., E.S., T.B.\ and 
A.S. were supported by the U.S.\ Dept.\ of Energy under contracts 
DE-AC02-98CH10886, and DE-FG02-92ER40716. J.N. was partially supported 
by the Japan Society for the Promotion of Science.  We thank 
Chris Sachrajda, and Steve Sharpe for useful discussions and the 
referee for questioning our original treatment of the chiral limit.

\bibliography{paper}

\end{document}